\documentclass[aps,amsfonts,onecolumn,a4paper,showpacs,showkeys]{revtex4}  

\begin{document}
\title{Boson-fermion stars: exploring different configurations}
\author{Alfredo B. Henriques} 
\affiliation{Departamento\ de F\'{\i}sica/CENTRA,
  Instituto Superior T\'ecnico, Av. Rovisco Pais, 1096 Lisboa, Portugal}
\email{alfredo@fisica.ist.utl.pt} 
\author{Lu\'{\i}s E. Mendes}
\affiliation{Astrophysics Group, Blackett Laboratory, Imperial College
  London SW7 2BW, U.~K.} \email{l.mendes@imperial.ac.uk} 
\date{\today}
\begin{abstract}
  We use the flexibility of the concept of a fermion-boson star to
  explore different configurations, ranging from objects of atomic
  size and masses of the order $10^{18}$ g, up to objects of galactic
  masses and gigantic halos around a smaller core, with possible
  interesting applications to astrophysics and cosmology, particularly
  in the context of dark matter.
\end{abstract}

\maketitle
\section{Introduction}
Scalar fields have been playing an increasingly important role in
recent investigations in cosmology and astrophysics. A number of such
fields have been considered, ranging from the dilaton in superstrings,
to the Higgs field and from the postulated inflaton, to the axion.
This is one of the reasons for the interest in the properties of
stellar objects made out of boson fields, called boson stars, since it
is quite possible that such objects play, or have played, a
considerable role as components of the dark matter of the
universe. Moreover, such exotic objects provide us with unique
insights into very high energy physics which, apart from the imprints
left in the Early 
Universe, is currently almost completely unaccessible to
experiment\cite{klop}. 

Boson stars can be seen as macroscopic boson condensates, formed by
the vacuum expectation value of scalar fields, the particles being the
excitations over the vacuum expectation value. First studied by Ruffini
and Bonnazola~\cite{rb69}, their work was later generalised to include
self-interactions~\cite{csw86}, gauge charges~\cite{jb89}, excited
solutions of the coupled Einstein and Klein-Gordon equations, as well
as non-topological soliton solutions~\cite{flp87,l87},
Q-stars~\cite{l89}, oscillating soliton stars~\cite{ss91}, and even
stars with two bosons with different masses~\cite{dh98}. The stability
of these stars was studied in~\cite{j89} and~\cite{ty84}.  Two
complete reviews on the subject can be seen in references~\cite{lm92}
and~\cite{j92}.

However, if these stars had their origin in some primordial gas of
both bosons and fermions, one would certainly expect to find objects
made out of a mixture of these two types of particles. Even if they
were purely bosonic or fermionic at their origin, they would
later be susceptible to a considerable amount of contamination by
bosons or fermions, respectively. This is why it is important to
investigate these mixed objects, now called boson-fermion stars. We
shall not study here the important question of how these stars, both
pure boson or boson fermion might
have been formed, by gravitational condensation, out of a primordial
gas. This issue has been addressed in part, for pure scalar fields,
in~\cite{kmz85} and~\cite{ml90}. We shall assume these mixed stellar
objects to exist and study the configurations they can take as stable
systems.

Their stability and main properties have been investigated
in~\cite{hlm90,hlm90_2,sts98}. Generalising simple heuristic arguments
by Thirring~\cite{t83}, it was then pointed out that many of the general
properties could be understood on the basis of an interplay between
the Pauli principle and the Heisenberg principle, the first being
ultimately responsible for the stability of the fermionic component
and the uncertainty principle being 
responsible for the stability of the bosonic component. The 
dependence on these two fundamental principles of quantum mechanics is
responsible for the very different characteristics shown by these
stars, when we have configurations that are either predominantly
fermionic or predominantly bosonic.

In most of the former investigations on the subject, the fermions were
taken to be neutrons. We now relax this assumption, taking the fermion
mass as a free parameter, accepting as reasonable that we may have different
varieties of fermions in a primordial gas.  As for the boson
field, we take it to be a real, massive, self-interacting scalar
field, with the mass of its associated particle also taken as a free parameter.

Now that new ideas, like the one provided by fermion
balls~\cite{mv02} and supermassive boson stars~\cite{tcl00}, are being
put forward to explain the properties of 
galaxies, particularly the possible existence of supermassive dark
objects at galactic centers, or of using boson stars in  the role of
gravitational 
lenses~\cite{ds00} it is appropriate to come back to the
question of fermion-boson stars and to investigate if they can provide
us with alternative scenarios of interest in astrophysics and
cosmology, in particular when coupled with the problem of dark matter.
As a result of the properties of bosons and fermions, we are able, for
instance, to construct compact fermionic cores, more or less massive,
depending on the mass taken by the fermion, surrounded by immense
halos of bosons, whose sizes are also dependent on the mass of the
corresponding boson. Very different objects can be constructed, by
varying the masses of their components.

Modern high-energy physics theories providing us with a large variety
of bosons and fermions as serious candidates for dark matter,
including massive sterile neutrinos, gravitinos, axinos as well as
Higgs fields, axions, heavy scalars resulting from non-thermal decays
of inflatons, and still others~\cite{e00}, make it worth while to
pursue these theoretical constructions and explorations.
\section{The model}
The equations have been derived before, nevertheless, and for
completeness, we repeat them here. The metric we shall
use is the one appropriate to a stationary, spherically symmetric
distribution of matter:
\begin{equation}
  \label{eq:metric}
  ds^{2} = -B(r) \, dt^{2} + A(r) \, dr^{2} + r^{2} d\theta^{2} + 
  r^{2} \, \sin^{2}\!\theta \, d\phi^{2};
\end{equation}

The Lagrangian for the massive real scalar field is defined by
\begin{equation}
  \label{eq:lagrangian}
  {\cal L} = -\frac{1}{2} g^{\mu\nu} \partial_{\mu} \varphi \,
  \partial_{\nu} \varphi - \frac{1}{2} m^{2}\varphi^{2} -
  \frac{1}{4} \lambda
  \varphi^{4}
\end{equation}
and from it we derive the energy-momentum tensor:
\begin{equation}
  \label{eq:enmom}
  T_{\mu\nu} = \partial_{\mu}\varphi \,
  \partial_{\nu}\varphi - g_{\mu\nu} \left( \frac{1}{2} g^{\rho\sigma} 
  \partial_{\rho}\varphi \, \partial_{\sigma}\varphi + \frac{1}{2}  m^{2} 
  \varphi^{2} + \frac{1}{4} \lambda \varphi^{4} \right)\,.
\end{equation}

We expand the scalar field in terms
of creation and annihilation operators in the usual
manner~\cite{ml90,hlm90,hlm90_2}, in such a way that $\varphi^{+} =
\varphi$, as is appropriate for a real field: 
\begin{equation}
  \label{eq:phiexp}
  \varphi(r,t)= \Sigma_{n} \frac{1}{(2\omega_n)^{1/2}}
  \left( a_{n} \varphi_{n}\!(r) e^{-\imath\omega_{n} t} +
  a_{n}^{+} \varphi_{n}^{+}\!(r) e^{\imath\omega t}\right) 
\end{equation}
with
\begin{equation}
  \label{eq:com}
  [a_{m},a_{n}^{+}] = \delta_{mn} \,.
\end{equation}
We shall assume in our problem all the bosons to be in the discrete
quantum state defined by $n=0$, defining the lowest energy of the
star, $\omega_{0}$. This corresponds to a spherically symmetric and
nodeless wave function $\varphi_{0}(r)$. It also means that the
bosons, in number $N_{B}$, will be in the quantum state
\begin{equation}
  \label{eq:quant_state}
  \arrowvert N_{B} \rangle = (N_{B}!)^{-1/2}
  (a_{0}^{+})^{N}_{B}\,\arrowvert 0 \rangle \,.
\end{equation}

Following Breit et al.~\cite{bgz84}, we find that the canonical
commutation relations require that
\begin{equation}
\label{eq:norm}
\int 4 \pi r^{2} \sqrt{\frac{A}{B}}\,
\arrowvert\varphi_{0}\arrowvert^{2} dr = 1 \,.
\end{equation}
The Einstein equations being classical equations, in order to write
their right hand side we take the following expectation value for the
energy-momentum tensor:
\begin{equation}
  \label{eq:enmom_class}
  T_{\mu\nu} = \langle N_{B} \arrowvert :\!T_{\mu\nu}\!: \arrowvert N_{B}
  \rangle=\partial_{\mu}\varphi_{c}^{+}
  \partial_{\nu}\varphi_{c} - \frac{1}{2} g_{\mu\nu} \left( g^{\rho\sigma} 
  \partial_{\rho}\varphi_{c}^{+} \partial_{\sigma}\varphi_{c} + m^{2}
  \varphi_{c}^{+}\varphi_{c} + \frac{1}{2} \lambda |\varphi_{c}|^{4}\right)
\end{equation}
where we introduced, with the help of eq.~(\ref{eq:quant_state}), the
semi-classical field
\begin{equation}
  \label{eq:classfield}
  \varphi_{c}(r,t)= (\omega_{0})^{1/2} (N_{B} + \frac{1}{2})^{1/2}
  \varphi_{0}(r) e^{-\imath \omega_{0} t}\,.
\end{equation}
The normalisation condition now takes the form
\begin{equation}
  \label{eq:normclass}
  \int 4 \pi r^{2} \sqrt{\frac{A}{B}} \, \varphi_{c}^{2} \, dr =
  \frac{1}{\omega_{0}} \left( N_{B} + \frac{1}{2} \right) \,.
\end{equation}

It is well known that Fermi statistics leads the energy-momentum
tensor to approach the perfect fluid form, when the
number of fermions present is large~\cite{rb69,ov39}. The equation we
shall use for the degenerate fermion gas was first derived by
Chandrasekhar and has the following parametric form:
\begin{eqnarray}
  \label{eq:paramEqSt}
  \rho & = & K (\sinh T - T) \\
  p    & = & \frac{K}{3}  (\sinh T - 8 \sinh \frac{T}{2} + 3 \, T) \, ,
\end{eqnarray}
the constant $K$ being given by $K = m_{f}^{4} / (32 \, \pi^{2})$ where
$m_{f}$ is the mass of the fermion, which we shall leave as a free
parameter. The parameter T is defined in terms of the maximum value of
the momentum of the Fermi distribution at radius $r$ by
\begin{equation}
  \label{eq:paramt}
  T(r) = 4 \log\!\!\left( \frac{p_{0}}{m_{f}} + 
  \sqrt{1 + \frac{p_{0}^2}{m_{f}^2}} \,\, \right) \,.
\end{equation}
Finally, the particle density $n(r)$ is given by the expression
\begin{equation}
\label{eq:density}
n (r) = \frac{p_{0}^{3}}{3 \, \pi^{2}} = \frac{m_{f}^{3}}{3 \, \pi^{2}}
\, \sinh^{3} \!\frac{t(r)}{4} \,.
\end{equation}

In the absence of an explicit non-gravitational interaction between
fermions and bosons, the Einstein equations are
\begin{equation}
\label{eq:einsEq}
R_{\mu\nu} - \frac{1}{2} g_{\mu\nu} R = 8 \pi G \left(
  T_{\mu\nu}(\varphi_{c}) +  T_{\mu\nu}(p,\rho) \right) \, ,
\end{equation}
with $T_{\mu\nu}(\varphi_{c})$ given by eq.~(\ref{eq:enmom_class})
and the fermion part is defined by the perfect fluid expression
\begin{equation}
  \label{eq:enMon_fluid}
  T_{\mu\nu}(p,\rho) = (\rho + p) u_{\mu} u_{\nu} + g_{\mu\nu} \, p
\end{equation} 
$u_{\mu}$ being the 4-velocity of the fluid. From the Bianchi
identities we find that the covariant derivative of the
energy-momentum tensor must be zero, from which condition we derive the
following two equations, which will then be coupled to the Einstein
equations:
\begin{equation}
\label{eq:enCond}
\frac{dp}{dr} = - \frac{1}{2} \frac{B^\prime}{B} (\rho +p )
\end{equation}
and, for the scalar field,
\begin{equation}
\label{eq:pCond}
\Box \varphi_{c} = m^{2} \varphi_{c} + \lambda \, \arrowvert
\varphi_{c}\arrowvert^{2} \varphi_{c} 
\end{equation}
where $\Box$ is the d'Alembertian operator associated with our metric.
Before we proceed, we must remember that the explicit time dependence
that appears in this equation and in eq.~(\ref{eq:enmom_class})
disappears, the exponential factor $\exp(-\imath \omega_{0} t)$ being
cancelled out; the result of the action of time derivatives is
then the term $\omega_{0}^{2} \varphi_{c}$.

We introduce some redefinitions of the variables~\cite{hlm90_2}, which
enable us to write the equations in a very convenient dimensionless
form ($c=\hbar=1$):
\begin{equation}
\label{eq:redef}
x=mr \, ; \;\; \sigma(x)=(4 \pi G)^{1/2} \varphi_{c}(r,0) \, ; \;\;
\Omega=\frac{\omega_{0}}{m} \, ; \;\; \Lambda=\frac{\lambda}{4 \pi G
  m^{2}} \, ; \;\; \bar\rho = \frac{4 \pi G \rho}{m^{2}} \, ; \;\; \bar p=
  \frac{4 \pi G p}{m^{2}} \,.
\end{equation}

The equations to be integrated become:
\begin{eqnarray}
\label{eq:bfsEq}
A^\prime & = & - \frac{A(A-1)}{x} + x A^{2}\left( \Omega^{2} \,
  \frac{\sigma^{2}}{B} + 
\frac{\sigma^{\prime 2}}{A} + \sigma^{2} + \frac{1}{2} \lambda \, \sigma^{4} +
2\bar \rho \right) \, ;\\
B^\prime & = & \frac{B(A-1)}{x} + x AB \left( \Omega^{2} \,
  \frac{\sigma^{2}}{B} + \frac{\sigma^{\prime 2}}{A} - \sigma^{2} -
  \frac{1}{2} \lambda \, \sigma^{4} + 
2\bar p \right) \, ;\\
\sigma^{\prime\prime} & = &  - \left( \frac{B^\prime}{2B} -
  \frac{A^\prime}{2A} + \frac{2}{x}\right) 
\sigma^\prime - \frac{A}{B} \, \Omega^{2} \sigma + A \sigma + A \lambda
  \, 
\sigma^{3} \, ;\\
T^\prime & = & -\left( \frac{d\bar p}{dT} \right)^{\!\!\!-1}
  \!\frac{B^\prime}{2B} \left( \bar \rho + \bar p \right) \, ,
\end{eqnarray}
where the last equation is derived from eq.~(\ref{eq:enCond}) and
\begin{eqnarray}
\label{eq:anotherEn}
\bar\rho & = & \alpha \left( \sinh T - T \right) \, , \\
\label{eq:anotherP}
\bar p & = & \frac{1}{3} \alpha \!\left( \sinh T - 8 \sinh\frac{T}{2}
  + \frac{T}{3} \right)
\end{eqnarray}
and $\alpha = m_{f}^{4}/(8 \pi m ^{2} m_{Pl}^{2})$, $m_{Pl}= G^{-1/2}$
being the Planck mass.

To calculate the total mass of the star we use
\begin{equation}
\label{eq:mass1}
A(x) = \left( 1 -2 \frac{M(x)}{x} \right)^{\!\!\!-1} \,.
\end{equation}
Integrating the equations up to a point sufficiently far out that we
may consider the fermion and boson terms to be negligible, we obtain the
mass from $x$
\begin{equation}
\label{eq:mass}
M(x) = \frac{x}{2} \left( 1 - \frac{1}{A} \right) \,.
\end{equation}

The number of fermions and the number of bosons in our star can also
be obtained from analytical expressions. From eqs.~(\ref{eq:normclass})
and~(\ref{eq:density}) we find
\begin{eqnarray}
\label{eq:numberB}
N_{B} & = & \omega \! \int \! 4 \pi r^{2} \sqrt{\frac{A}{B}} \,
\varphi_{c}^{2} \, 
dr = \Omega \left( \frac{m_{Pl}}{m} \right)^{2} \int \!
\sqrt{\frac{A}{B}} \, \sigma^{2} x^{2} \, dx \\[2mm]
\label{eq:numberF}
N_{f} & = & \int \! 4 \pi r^{2} \sqrt{A} \, n \, dr = \frac{4}{3 \pi}
\left( \frac{m_{f}}{m} \right)^{3} \! \int \! \sinh^{3}\! \frac{T}{4}
\, \sqrt{A} \, x^{2} dx \, \,,
\end{eqnarray}
where we have written both the original expressions and the ones
obtained after the redefinitions introduced above. These values enter
into the calculation of the binding energy, defined by
\begin{equation}
  \label{eq:bindingEn}
  E_{B}=M - \left( m_{b} N_{b} + m_{f} N_{f} \right) \,.
\end{equation}
A positive value of $E_{B}$ means an unstable configuration. For a
more complete discussion of the issue of stability in boson-fermion
stars we refer the reader to~\cite{hlm90}.

Although we cannot separate the mass of the fermionic
component of the star, we introduce in addition, as a useful approximation, but
only as an approximation, the following quantity for the fermion core:
\begin{equation}
\label{eq:coreMass}
M_{f} = \int 4 \pi r^{2} \rho dr \,.
\end{equation}

From the structure of the equations, we find that the initial
conditions are subjected to the following restrictions: $A(0)=1$ and
$\sigma^\prime(0)=0$, with $\sigma(0)$ and $T(0)$ free parameters. There is no
restriction on $B$, except that $B=1$ at infinity. As the equations are
linear in $B$, this can always be obtained by rescaling $B$. Apart from $B$,
only $\Omega$ needs to be rescaled.

Using the redefined variables, once we fix the
initial values $\varphi_{c}(0)$ and $T(0)$, and the value of $\alpha$,
our equations and their solutions are mathematically the same, for any
combination of the fermion and boson masses giving the same $\alpha$
in eqs.~(\ref{eq:anotherEn}) and~(\ref{eq:anotherP}). Only the scales of
the problem will be changed. From eq.~(\ref{eq:redef}) 
we see that, putting $\sigma(0)=1$, $\varphi_{c}(0)$ is close to the
Planck energy.

In the numerical simulations that follow we put 
$\lambda=0$. The dynamical influence of this constant on the total
mass of the star has been explored in
references~\cite{csw86},~\cite{ml90} 
and~\cite{hlm90_2}, but 
such influence is not particularly relevant for our considerations
here. We could also have introduced direct couplings between the
fermion and the boson fields, either in the spirit it has been done
in~\cite{sts98} and~\cite{bmh93} or, in a more phenomenological way,
in~\cite{ekow89}. They lead 
to limits in the couplings and masses of the scalar particles. We made
sure that, using values within the limits set up 
in~\cite{ekow89}, no important alterations would came to our results. 
\section{Numerical simulations and discussion}
From the equation for the scalar field, comparing the mass
term with the term involving the eigenvalue, corresponding to the
second time derivative of $\varphi$, we would in general expect the
eigenvalue $\omega$ to be approximately of the same order of magnitude as the
mass of the scalar field. This is confirmed by our numerical
simulations. The eigenvalue $\omega$ appears as a factor in front of
eq.~(\ref{eq:numberB}) for the number of scalars. In most
simulations, the mass of the scalar field is indeed very small. No 
wonder then that, to obtain configurations with a reasonable boson
population, we find it necessary to use at the origin values for
$\varphi_{c}(0)$ which are close to the Planck value.

For the fermions the situation is different. Using a value of
$T(0)$ around $1$, and the neutron mass ($m_n$) for the fermion, we find
parameters typical of neutron stars, far away from the Planck scale. When
we change $m_{f}$, these values are obviously scaled accordingly.

All we have to ask now is how reasonable is it to take such
large values for $\varphi_{c}(0)$ in the simulations. If we assume
such stars to have formed in the early universe, as usually
is the case when we deal with objects like 
primordial black holes, then planckian-like values seem reasonable.
When the stars are formed at later stages, this assumption is more
questionable. It is thus important that we do not restrict our
attention to a too small domain of values for $\varphi_{c}(0)$.
Likewise, we shall explore a large domain of masses for the fermions
and the bosons, not restricting ourselves to known or expected masses.

We shall apply an iterative shooting method to determine the
eigenvalue $\omega$. A technical problem appears when we deal with a
combination of 
masses such that the boson wave function quickly decreases to very
small values, well inside the fermion core. We have to make sure this
is not due to an overshooting. The way to do it is to make sure that,
by displacing the node of $\sigma(x)$, there are no changes in the
total mass of the star and in the number of bosons, which in turn
requires that the eigenvalue be fixed with great
accuracy~\cite{hlm90_2}. In what follows we shall study two different
situations which differ mainly by the relative magnitude of the bosonic
and fermionic terms in the Einstein equations.
\subsection{Fermionic and bosonic terms of the same order of
  magnitude}
To set the problem, we begin by what looks like a natural
choice, taking the fermion mass equal to the neutron mass and assuming
the different terms in our equations to be of the same order of
magnitude. This requires a $\varphi_{c}(0)$ not very distant from
the Planck scale. We will now analyse four different cases:\newline
\textbf{Case 1:} $T(0)=1 \,;\,\,\, \varphi_{c}(0) = 2.3\times10^{17}
 \textrm{GeV}\, (\sigma(0)=0.1) \,;\,\,\, m_{f}=m_n \,;\,\,\,
 m_{b}=2.71\times10^{-20} \textrm{GeV}$
\par In this case we find
\begin{eqnarray*}
\rho(0) & = & 9.96\times10^{13} \,\textrm{g\,cm}^{-3} \,; \\
M & = & 4.76\times10^{33} \,\textrm{g} \approx M_\odot
\,\,\,(M_{f}=3.44\times10^{32}\, \textrm{g} \approx 0.1 M_\odot) \,;\\
N_{b} & = & 9.32\times10^{76} \,\,\,(N_{b} m_{b} = 4.50\times10^{33}\,
\textrm{g} \approx M_\odot) \,;\\
N_{f} & = & 2.08\times10^{56} \,\,\,(N_{f} m_{f} =3.47\times10^{32}\,
\textrm{g} \approx 0.1  M_\odot)\,;\\
E_B/c^2 & = & - 9.0\times10^{31}\,\textrm{g}\,;\\ 
r_{c} & = & 15.9 \,\textrm{km} \,;\\
r_{90}& = & 68.9 \,\textrm{km} \, ;
\end{eqnarray*}
where $\rho(0)$ is the central density corresponding to the fermionic
component, $r_{c }$ is the radius of the fermion core and $r_{90}$ the
radius within which $90\%$ of the total mass of the star is
concentrated. This is a value we shall always quote whenever, as is
the present case, the tail of the wave function of the scalar field
extends beyond the fermion core, forming a halo around it. From the
inspection of $N_{b} m_{b}$ and $N_{f} m_{f}$, or $M$ and $M_{f}$, we
conclude that the present configuration is dominated by the boson
component which accounts for more than $90\%$ of the total mass. This is
an example of values we have found in ref.~\cite{hlm90_2}. The binding
energy was converted into the mass equivalent.

Keeping the same fermion mass, we go to higher boson masses by
giving smaller values to $\alpha$. For instance, with
$\alpha=10^{-24}$, we find $m_{b}=2.7\times10^{-8}$ GeV and we have to
decide what is a reasonable choice for $\varphi_{c}(0)$. One
possibility is to assume that, once the gravitational field couples to
the energy, the terms in the equations are of the same order of
magnitude, which was the philosophy we used above, except that this
time this requires a value for $\varphi_{c}(0)\approx10^{6}$ GeV. It
is instructive to compare this 
with the case where we give to the scalar field a value close to the
Planck scale. First the case where the terms are about equal.\newline
\textbf{Case 2:} $T(0)=1 \,;\,\,\, \varphi_{c}(0) = 2.3\times10^{6}
\textrm{GeV} \,(\sigma(0)\approx10^{-12}) \,;\,\,\, m_{f}=m_n
\,;\,\,\, m_{b}=2.71\times10^{-8} 
\textrm{GeV}$
\par Now we get:
\begin{eqnarray*}
\rho(0) & = & 9.96\times10^{13} \,\textrm{g\,cm}^{-3} \,; \\
M & = & 6\times10^{32} \,\textrm{g} \approx 0.1 M_\odot\,\,\,
(M_{f}\approx M) \,; \\ 
N_{b} & = & 3.70\times 10^{47} \,\,\, (N_{b}m_{b}=1.79\times10^{16}\,
\textrm{g} \approx 10^{-17} M_\odot) \,;\\
N_{f} & = & 3.62\times10^{56}\,\,\, (N_{f}m_{f}=6.04\times10^{32}\,
\textrm{g} \approx 0.1 M_\odot) \,;\\
E_B/c^2 & = & - 5.44\times10^{30}\,\textrm{g}\,;\\ 
r_{c} & = & 20.8 \,\textrm{km}\,;
\end{eqnarray*}
where $r_{90}$ is not quoted, as now the scalar wave function goes
practically to zero well inside the fermion radius. The star is
entirely fermion dominated and it has the corresponding typical
physical quantities, like mass and radius. If we increase $T(0)$, the
number of fermions increases, as expected, but the number of bosons
remains practically the same. This is a typical configuration to be
considered when dealing with neutron stars contaminated by WIMPS. If
the masses of the WIMPS are of the order of $10^{-5}$ eV or above,
they will concentrate in the core of the star.

In the following example, the scalar field is closer to the
Planck scale:\newline
\textbf{Case 3:} $T(0)=1 \,;\,\,\, \varphi_{c}(0) = 2.3\times10^{17}
\textrm{GeV} \,(\sigma(0)=0.1) \,;\,\,\, m_{f}=m_n  \,;\,\,\, m_{b}
= 2.71\times10^{-8} \textrm{GeV}$
\par Now we find 
\begin{eqnarray*}
\rho(0) & = & 9.96\times10^{13} \,\textrm{g\,cm}^{-3}\,; \\
M & = & 5.23\times10^{21} \,\textrm{g}  = 10^{-12} M_\odot\,\,\,
(M_{f}=3.14\times10^{-3}\, 
\textrm{g} ) \,;\\
N_{b} & = & 1.11\times10^{53} \,\,\, (N_{b}m_{b} = 5.33\times10^{21}\,
\textrm{g} \approx 10^{-12} M_\odot) \,;\\
N_{f} & = & 1.89\times10^{21} \,\,\, (N_{f}m_{f}=3.17\times10^{-3}\,
\textrm{g})\,;\\
E_B/c^2 & = & - 9.99\times10^{19}\,\textrm{g}\,;\\ 
r_{c} & = & 3.13\times10^{-6} \,\textrm{cm} \,;\\
r_{90} & = & 7.0\times10^{-6} \,\textrm{cm} \,.
\end{eqnarray*}
We have a configuration which is entirely different from the
preceding case, although we have kept the same central density
$\rho(0)$. It is now a typical soft boson star, with a microscopic
radius, but a still respectable mass. The fermion component has almost
completely disappeared. The increase in $\varphi_{c}(0)$ has
resulted, not only in a substantial increase in $N_{b}$, but also in
an enormous reduction in the number of fermions. No wonder this has
happened as, with the choice made for $\varphi_{c}(0)$, the terms in
the equations associated with the bosons are very much larger than the
terms associated with the fermions. In this sense, it is not a natural
choice.

If we increase $T(0)$ we find that, around $T(0)=3$, the
configuration will suddenly become fermion dominated, a property that
has been noticed and investigated in the work of
reference~\cite{hlm90_2}, and we 
shall have a configuration similar to the one in case 2 above, albeit
with a very different value for $\phi_c(0)$.

Before going to a different kind of limit, we mention one case
corresponding to the fermions and bosons having much higher masses
than the ones used so far:\newline
\textbf{Case 4:} $T(0)=1 \,;\,\,\, \varphi_{c}(0) =
2.3\times10^{6}\textrm{GeV}  \,(\sigma(0)\approx10^{-12}) \,;\,\,\,
m_{f}=10^{7}\textrm{GeV}  \,;\,\,\,  m_{b}=3.08\times10^{6}\,
\textrm{GeV}$
\par In this case:
\begin{eqnarray*}
\rho(0) & = & 1.29\times10^{42} \,\textrm{g\,cm}^{-3} \,; \\ 
M & = & 5.28\times10^{18} \,\textrm{g} \approx 10^{-15} M_\odot\,\,\,
(M_{f}\approx M) \,;\\ 
N_{b} & = & 2.90\times10^{19} \,\,\, (N_{b} m_{b} = 1.59\times10^{2}\,
\textrm{g}) \,;\\
N_{f} &= & 2.99\times10^{35} \,\,\, (N_{f}m_{f}=5.33\times10^{18}\,
\textrm{g})\,;\\
E_B/c^2 & = & - 4.79\times10^{16}\,\textrm{g}\,;\\ 
r_{c} & = & 1.83\times10^{-8} \,\textrm{cm}
\end{eqnarray*}
It is really an object of atomic size, with characteristics close to
primordial black holes, without being one. The value we took for
$\varphi_{c}(0)$ is within the same order of magnitude as the masses
and is a case where the fermion and boson terms in the equations are
of the same order of magnitude.

Again, it is completely fermion dominated. Except for the case
1 above, in all the others, whenever the terms associated with the
bosons and the fermions are of about the same order of magnitude, the
resulting configuration is dominated by fermions. This happens, for
$m_{f}=m_{n}$, when $m_{b}$ is larger than $10^{-5}$ eV, as in case 2 above.
\subsection{Fermionic terms dominate}
We shall now investigate what happens when $m_{f}$ is equal
to or smaller than the nucleon mass and the boson mass is really very
small. This means that we have to increase $\alpha$. The fermion terms
will necessarily dominate the equations, whatever value we use for
$\varphi_{c}(0)$, if we keep it, as we shall do, equal or below the
Planck scale. However, due to the smallness of the boson mass, this
does not mean that the star will be fermion dominated. In case of pure
boson stars, we know that the total mass of the star tends to change
according to $1/m_{b}^{2}$. As a first example we take:\newline
\textbf{Case 5:} $T(0)=1 \,;\,\,\,
\varphi_{c}(0)=2.3\times10^{17}\textrm{GeV}   \,(\sigma(0)=0.1)
\,;\,\,\,m_{f}=m_n \,;\,\,\,   
m_{b}=8.56\times10^{-23}\textrm{GeV}$
\par Now we have
\begin{eqnarray*}
\rho(0) & = & 9.96\times10^{13} \,\textrm{g\,cm}^{-3} \,; \\
M & = & 1.65\times10^{36} \,\textrm{g} \approx 10^{3} M_\odot
(M_{f}=5.99\times10^{32}\textrm{g} \approx 0.1 M_\odot) \,;\\
N_{b} & = & 1.10\times10^{82} \,\,\, (N_{b}m_{b}=1.68\times10^{36}\,
\textrm{g} \approx 10^{3} M_\odot) \,;\\
N_{f} & = & 3.61\times10^{56} \,\,\,  (N_{f}m_{f}=6.04\times10^{32}\,
\textrm{g} \approx 0.1 M_\odot)\,;\\
E_B/c^2 & = & - 3.16\times10^{34}\,\textrm{g}\,;\\ 
r_{c} & = & 20.8 \,\textrm{km}\,;\\
r_{90} & = & 2.18\times10^{4}  \,\textrm{km}\, \;.
\end{eqnarray*}
We have a complicated configuration, with a core typical of a neutron
star, and a very massive boson halo extending well beyond. The bosons
dominate the total mass by many orders of magnitude.

Had we used a still smaller boson mass, for instance $m_b =
2.71\times10^{-32}$ GeV, we would have got a star with a total mass of
$M = 5.26\times10^{45}\,\textrm{g}\approx 10^{12} M_\odot$ and a halo
extending up to $6.92\times10^{13}\,\textrm{km} = 2.24\,\textrm{pc}$,
most of the mass coming from the boson component (the approximate
value for the fermion component is $M_f =
5.99\times10^{32}\,\textrm{g}\approx 0.5 M_\odot$ and $r_c = 20.8$ km,
the same as above).
 
The authors of~\cite{mv02} have tried to show that, choosing a fermion
mass of about $16$ KeV (for instance, a sterile neutrino), it was possible
to explain the main dynamical features of the supermassive compact
dark objects, known to exist in the Galactic center and in the center
of M87, with the help of what they have dubbed fermion balls, which
they propose as an alternative explanation to the more usual black
hole model. Better observations~\cite{s02} have, however, ruled out
this model, but still leaving opened the possibility of a supermassive
boson star~\cite{tcl00}.

These ideas suggest that we consider a boson-fermion star with both a
small fermion mass and a very small boson mass. When we consider this
possibility we get the following configuration:\newline
\textbf{Case 6:} $T(0)=1 \,;\,\,\,
\varphi_{c}(0)=2.3\times10^{17}\textrm{GeV}  \,(\sigma(0)\approx0.1)
\,;\,\,\, 
m_{f}=10^{-5} \textrm{GeV}   \,;\,\,\,  m_{b}=3.08\times10^{-30}
\textrm{GeV}$
\par For this last case we have
\begin{eqnarray*}
\rho(0) & = & 1.29\times10^{-6} \,\textrm{g\,cm}^{-3} \,; \\
M & = & 4.18\times10^{43}  \,\textrm{g} \approx 10^{10} M_\odot
\,\,\, (M_{f} = 3.03\times10^{42}  \,\textrm{g} \approx 10^{9} M_\odot ) \,;\\ 
N_{b} & = & 7.22\times10^{96} \,\,\,  (N_{b} m_{b}=3.96\times10^{43}\,
\textrm{g} \approx 10^{10} M_\odot) \,;\\
N_{f} & = & 1.71\times10^{71} \,\,\,  (N_{f}m_{f}=3.05\times10^{42}\,
\textrm{g} \approx 10^{9} M_\odot) \,;\\
E_B/c^2 & = & - 7.98\times10^{41}\,\textrm{g}\,;\\ 
r_{c} & = & 1.40\times10^{11} \,\textrm{km}\,=4.53\times10^{-3}\textrm{pc} \\
r_{90} & = & 6.09\times10^{11}  \,\textrm{km}\,=1.97\times10^{-2}
\,\textrm{pc} \\  
\end{eqnarray*}

The fermion core and the boson halo form a supermassive
compact object with the halo extending up to $2\times10^{-2}$
pc. Keeping the same fermion mass, but taking, for 
instance, a boson mass in the range of $10^{-32}$ GeV, we would get a
total mass of the order of $10^{13}$ solar masses, with a halo
extending up to $6$ pc, while the fermion component would have
practically the same mass and the same radius as in case 6.

Would it in principle be possible to distinguish a
boson-fermion system from a fermion ball? Once again this could be
done through the careful investigation of the orbits of stars around a 
candidate object. We must
take into account that the mass density profile of a boson-fermion
configuration, with an extended boson halo, would be different from
the profile of a fermion ball. This should be reflected in the
dynamics of the orbits of stars inside such a halo, exactly as there
appears a difference between the fermion ball and the black hole
models~\cite{mv02}. We are currently investigating this possibility
and the results will be reported in a separate publication~\cite{next}.

We are thus in the presence of a system which can take
many different configurations, characterised by very different masses
and radii, ranging from objects of atomic size and masses of the order
of $10^{18}$ g, to objects having galactic masses and extending up to a
few light years. To show this flexibility was the main purpose of the
present work.
\acknowledgments{The authors would like to acknowledge useful
  discussions with and suggestions from Domingos Barbosa, Andrew
  Liddle, Ilidio Lopes, Jo\~ao Seixas and Luis Ure\~na-Lopez. 

\end{document}